# Proposing LT based Search in PDM Systems for Better Information Retrieval

Zeeshan Ahmed

University of Wuerzburg Germany
Vienna University of Technology Austria

**Abstract**

PDM Systems contain and manage heavy amount of data but the search mechanism of most of the systems is not intelligent which can process user's natural language based queries to extract desired information. Currently available search mechanisms in almost all of the PDM systems are not very efficient and based on old ways of searching information by entering the relevant information to the respective fields of search forms to find out some specific information from attached repositories. Targeting this issue, a thorough research was conducted in fields of PDM Systems and Language Technology. Concerning the PDM System, conducted research provides the information about PDM and PDM Systems in detail. Concerning the field of Language Technology, helps in implementing a search mechanism for PDM Systems to search user's needed information by analyzing user's natural language based requests. The accomplished goal of this research was to support the field of PDM with a new proposition of a conceptual model for the implementation of natural language based search. The proposed conceptual model is successfully designed and partially implementation in the form of a prototype. Describing the proposition in detail the main concept, implementation designs and developed prototype of proposed approach is discussed in this paper. Implemented prototype is compared with respective functions of existing PDM systems .i.e., Windchill and CIM to evaluate its effectiveness against targeted challenges.

## 1. Introduction

In early 1970s there was no such system to automate the process of data management, then in 1980s Computer Integrated Manufacturing was introduced but seemed not to be successful in product data management. With emergence of CAD technologies PDM Systems are introduced and used to manage engineering data, activities and processes through better control of engineering data, activities, changes and product configurations. PDM products mainly manage information about design and manufacturing of products including technical operations and running projects. PDM is also renowned as Engineering Data Management and Engineering Document Management Systems because it provides better management and control over engineering data, activities, and changes related to design and manufacture of product. Product Lifecycle Management (PLM) is another acronym for PDM to manage the entire development life cycle of the product by integrating people, data, processes and business systems.

PDM provides a backbone for the controlled flow of engineering information throughout the product life cycle by using engineering data, such as CAD, ERP and field service. Moreover PDM also supports product teams and techniques by providing Concurrent Engineering in improving engineering workflow. PDM systems address issues such as control, quality, reuse, security and availability of engineering data. PDM performs five main functions to integrate and manage all applications, information, and processes during the associated product life cycle i.e., Data vault and document management, Workflow and process management, Product structure management, Parts management and Program management [8].

The major objectives of PDM are to reduce the cost of engineering, reduce effort in product development life cycle, reduce time in change handling and new product development, improve the quality and services of the product, deliver and support products at the given time, improve team coordination, increase customization of products, maintain product configuration based information, manage large volumes of data generated by computer based systems, reduce engineering environment based problems, provide better access to information, provide better reuse of design information, provide common data warehousing, secure engineering data's originality, prevent error creation and propagation and make a strong effect on market shares. Moreover PDM is also supposed to handle business process work flows, change management, revision control, product configurations, product structure management, project tracking and resource planning.

To meet the aforementioned objectives of PDM, the concepts turned into the real time applications called PDM systems. These systems are developed to manage product data throughout enterprise, ensuring the availability of right information for the right person at the right time and in the right form [9]. PDM systems are mainly used by project managers, designers, engineers, administrators, manufacturing, sales, marketing, purchasing and other personal in the companies. Product related information controlled by PDM systems includes part definitions and other design data, engineering drawings, project plans, software components of products, product specifications, NC programs, analysis results, correspondence, bills of materials etc. Commercial PDM systems have been developed and are used in companies for more than a decade now. Many companies today have realized strategic importance of a PDM system implementation and usage. But the implementations have often been associated with problems and large costs for the companies. Still there is lot of work to be done in order to improve PDM systems functionality and to develop methods for their proper implementation and use in different areas of the product development and the sales delivery process.

PDM Systems plays an important role in tracking products among different engineering groups by reducing time to market, increasing product quality and reducing total cost. Furthermore PDM System controls, manages and distributes product data automatically to the needed people. A PDM system is typically used within enterprise to organization to access and control data related to its products and to manage the life cycles of those products. PDM Systems are capable of providing user directed and utility functions. User directed functions are i.e. Data vault and document management for storage and retrieval of product information, workflow and process management procedures for handling product data and providing of a mechanism to drive a business with information, Product structure management handling of bills of material, product configurations, and associated versions and design variants, Parts management providing of information on standard components and facilitating reuse of designs. Program management provides work breakdown structures and allows coordination between product related processes, resource scheduling and project tracking. Where as the utility functions are the Communication and notification capabilities such as links to email provide support for information transfer and events notification, Data transport tracking of data locations and moving of the data from one location or application to another, Data translation file exchange in the proper format. Image services, storage, access, viewing and markup of product information, System administration system control and monitoring of operation and security.

In this paper; in section 2 a clear vision to the targeted problems is provided, then in section 3 related research work is discussed, section 4 presents the over all proposed approach where as section 5 presents the solution towards the targeted and discussed problem. Section 6 discusses implementation designs of proposed approach to develop it into the form of software application, discussed in section 7. Proving the effectiveness against targeted problem, implemented prototype is compared with some existing PDM Systems in section 8. Discussion is concluded in section 10 after the presentation of some existing limitations in prototype development in section 9.

## 2. Problem Definition

PDM Systems contain and manage heavy amount of data, which itself is a big achievement but on the other hand the problem starts when user needs to find out some information out of this heavy data. The search mechanism of most of the available PDM Systems is limited because it provides limited structured options to find out the information. If user needed information is available amongst those provided search options then it is fine but in case user needs some information which can't be found using provided options then the outcome will be limited. Moreover it also consumes time that at first the user needs to read the given search options, then to provide the needed information by filling forms and then performing search.

No doubt available search mechanisms of PDM Systems capable of searching results with high probability but it is time consuming, heavily structured, static and limited. There are some serious problematic issues in search mechanisms of existing PDM systems which are needed to be resolved. The search mechanism of most of the existing PDM systems is;

1. Based on a static way of searching information like filling forms and making search.
2. Not capable of processing user's structured / unstructured natural language based queries to search information.

3. Not capable of retrieving information by extracting Meta data out of data.
4. Not capable of providing geometrical search for graphical documents e.g. a user is interested in finding some CAD documents and he wants to make the search with some geometrical information like size of screw etc. but using existing PDM System's search mechanism it's not possible.
5. Not capable of performing system spanning based search e.g. a user is interested in finding some information needed to design a CAD document but that is not available in the default System, then it will not look for other web based available relevant information sources for CAD document designing. In short the existing PDM systems do not provide the multiple database connectivity and search; it only looks in its own connected database for the retrieval of the required information.
6. Not capable of weighting extracted results like some other search engines e.g. Google etc.

Currently available search mechanisms of some of the PDM Systems e.g. Windchill [1] as shown below in Fig. 1 and CIM Data Base [2] as shown below in Fig. 2 etc. are limited because they provide restricted options to find out the information.

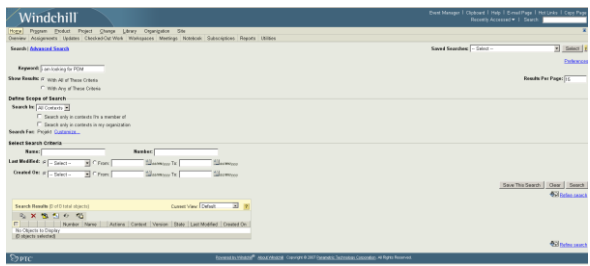

Fig. 1. Windchill Advanced Search Form

As shown in Fig.1, the presented advanced search mechanism of Windchill consists of a form with several different options to make search e.g. if a user is looking for a documents then one is required to enter the following information like name, number, last modified date etc and only then he can look for the required information. Windchill offers customization of the search form e.g. a user can enhance search form by adding or deleting some options from the main option list and can improve the search mechanism. Though Windchill is providing an efficient form based search mechanism, at the same time it is restricting the user to the existing search with provided options, consuming user's time in filling a form for finding some information etc. This Windchill search form also offers an option to make full text search but that can only work for one keyword e.g. "CAD". It does not allow to make search by entering natural language based short conditional objective statements e.g. "want CAD" or simple conditional objective statements e.g. "I am looking for CAD" or multiple conditions based objective statements e.g. "I am looking for CAD where document type is doc".

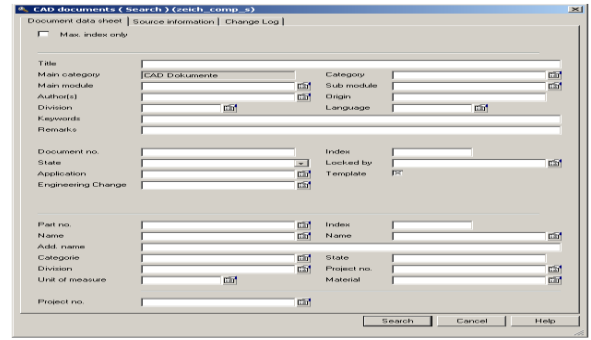

Fig. 2. CIM Database CAD Document Search Form

As shown in Fig. 2, the presented search form to search CAD documents using CIM Database consists of several different options like Title, Category, Author, Origin, Language etc. Like Windchill the search mechanism of CIM Database is also limited and time consuming. In both the cases (CIM and Windchill) user at first needs to fill some form to make search. Moreover both the systems are incapable of providing intelligent search, Meta data based full text search, geometrical search and multiple database spanning search. Moreover another deficiency in CIM Database search mechanism is that it is case sensitive by default. To make case insensitive search a user at first has to locate and change options. Like Windchill, CIM Database search forms also offers an option to make full text search but that can only work for one keyword e.g. "PDM". It does not allow to make search by entering natural language based short conditional objective statements e.g. "need PDM" or simple conditional objective statements e.g. "I am looking for PDM" or multiple conditions based objective statements e.g. "I am looking for PDM where document type is doc and pdf".

## 3. Related Research Work - Language Technology

Targeting the objective of this research; introduction and implementation of a new way for the implementation of a natural language based search mechanism to extract desired results from database by processing and modeling natural language based user requests, a thorough is conducted. Meeting the goals of this research, the filed of Language Technology is selected and explored.

Language Technology is a linguistic based field of computer science, also called as Human Language Technology or Natural Language Processing (NLP) [3]. Language technology is about to make machine capable of reading, listening, understanding and analyzing human (natural) language. The main objective of Language technology is to teach machines, how to communicate and help humans by communicating (listening and speaking) with them [4]. To meet aforementioned goal of natural language processing and make machines understand natural language, language technology is composed of two steps i.e., Tokenization and Parsing. Tokenization is also called lexical analysis, during the process of lexical analysis, natural language based instruction tokenizes in possible number of tokens by a lexer. Then these tokens are matched with the dictionary of used natural language for processing to identify valid and invalid tokens. These tokens consist of letters, digits and symbols. Then at the end of the process of lexical analysis a stream of tokens is generated by lexer for further processing. Generated token stream by lexer, then is considered by the parser to semantically evaluate the instruction. Every parser of any language programming or natural language consists of a set of rules. These set of rules are the combinations of tokens of dictionary of the language. To evaluate the semantic of a statement it is compulsory to first evaluate the tokens from dictionary and then the combinations of used valid tokens to understand the meaning or semantic of statement. To meet parsing goals, parsers are divided in to two types i.e., LL Parser and LR Parser. The LL parser constructs left most derivation and LR constructs eight most derivation during parsing, in simple words LL parser starts parsing by replacing nonterminals from left side and LR from right side. Moreover parser creates the sequences of tokens to put them into an Abstract Syntax Tree (AST).

In the domain of language technology, following the concepts of language technology, many approaches have been introduced by many researchers which are providing lots of values in the implementation of natural language processing i.e., Analyzing English Grammar [5], Layerd Domain Class [6] and Another tool for language recognition [7] etc.

*3.1. Analyzing English Grammar[5]*

Author has discussed an approach to analyze English grammar. Approach is divided into following three steps i.e., Division of grammar, Tokenization of Sentences using Lexer, Seven Steps Structuring and Categorization.

1. Division of grammar; Grammar is divided into two inter related studies: Morphology and Syntax. Morphology is to form the words in smaller units called morphemes, for example, the word "books" here would have two morphemes (i) the root/stem "book", and (ii) the inflectional morpheme {s} showing number [+Plural]. Syntax is to string words together to form (Partial) Phrases, Clauses, and (full) Sentences. For example, as presented above, the Determiner Phrase (DP) is formed from out of the string D+N.

2. Tokenization of Sentences using Lexer; Tokenizing sentences is by categorizing in two categories i.e., Class words and Functionals. Class words include Nouns, Verbs, Adjectives, and Adverbs while functional includes Determiners, Auxiliary/Modals, Pronouns, Complementizers, and Qualifiers.

3. Structuring and Categorization, as shown in Fig.3. To further structure the tokens of input sentences, a seven step guide is followed by author.

   a. Determining (DP) Articles, a/the; Demonstratives, this/that/these/those; Genitives my/our/your/their, etc.) precede Nouns: e.g., The book
   b. Determining Adjectives (AdjP) (modifiers of Nouns e.g., red, good, fast, etc.) precede and generally describe nouns [(Det)+Adj+N] (e.g., (The) read shoes
   c. Determining Main Verbs (MVP) (Tensed Verbs such as goes/went, walks/walked, keeps/kept, etc.) typically follow the subject of declarative sentences (adhering to the English SVO Subject Verb Object word order).
   d. Determining Auxiliary/Modals (AuxP) serve to introduce Main Verbs (MVPs). All functional features associated with Verbs {Tense, and Agreement features of Person/Number} are borne out of the Aux
   e. Determining Verb Phrase (VP) (Infinitive Phrase) unlike the MVP is a Non Tensed Verb Phrase. Such VPs tend to project after an already positioned MVP. These phrases include all three Infinitive types/forms e.g., I like to cook (=Infinitive 'to'), I like cooking (Infinitive 'ing'), I can cook (Infinitive 'bare verb stem' ).

| | |
|---|---|
| **Articles:** | a/an, the |
| **Demonstratives:** | this, that, these, those |
| **Possessives:** | my, your, his, her, its, our, their |
| **Indefinites:** | some, any, no, every, other, another, many, more, most, enough, few, less, much, either, neither, several, all, both, each, |
| **Cardinal Numbers:** | one, two, three, four,... |
| **Ordinal Numbers:** | first, second, third,...last |
| **Definition**: A Determiner is a functional structure-class word that precedes and modifies a Noun. | |
| **Features:** Definiteness, Case, Person, Number, (Gender) **Phrase Structure:** D + N → DP | |

Fig. 3. Summary of Determiners [5]

f. Determining Adverb Phrase (AdvP), like adjectives for nouns, modifies verbs e.g, softly touched, quickly ran, etc., (Adv+V).

g. Determining SVO/Head Initial Phrase: In addition to English being an SVO word order, English stipulates that the Head of a Phrase must be in the first initial position within the phrase (i.e., that word which labels the phrase such as Determiner, Adjective, Main Verb, etc. must come first in forming the phrase.

Analyzing English Grammar is highly relevant to the goal of this research. In this natural language processing module, the same mechanism is used to parse the sentences like tokenizing sentences, identifying relevant and irrelevant tokens with respect to the used grammar and then trying to analyze semantics of the input sentence.

*3.2. Layered Domain Class (LDC) [6]*

Authors developed software called Layer Domain Class (LDC) for parsing and deep semantic processing of English language based sentences. LDC consists of two major components and an external retrieval module. The first component, which was called "Prep," obtains information about a new domain and the language to be used in discussing it. The second, "user phase," component of LDC resembles an ordinary NL processor. To process English languages based sentences the whole process of parsing in LDC is divided into following four steps i.e., Scanning, Parsing, Semantic Processing and Output Generation as shown below in Fig. 4.

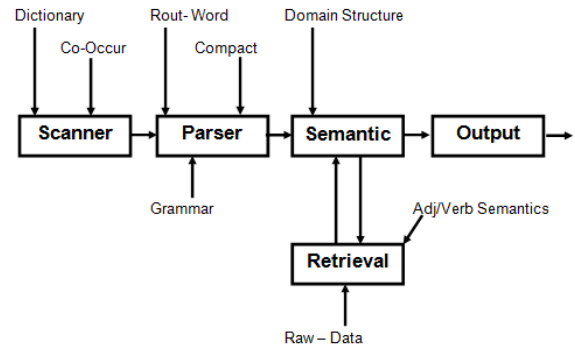

Fig. 4. Process of Phrase Parsing in LDC [6]

1. Scanner is to identify each word of the typed or spoken input and retrieve information about it from the dictionary file, which will have been created by Prep.
2. Parser is to determine, from the information provided by the scanner, the syntactic structure of the input. In a computational domain, especially one for retrieval rather than programming, the syntactic complexity of most inputs lies in the complexity of their noun phrases. For this reason relative clause verb forms are considered as basic, and sentence level verbs as derived.
3. Semantics module is to translate the tree like parse structures into an internal form that is referred to as "bubble structures". These structures, which can be interpreted directly or can be translated into a formal query for the external retrieval component, possess at least three user desirable properties.
4. Output generator converts the top level datarep produced by semantics into a human readable form.

*3.3. Another Tool for Language Recognition (ANTLR) [7]*

ANTLR is a tool, developed in 1983 by Professor Terence Parr and his colleagues to write grammar of languages, although this tool is only used for writing programming languages grammars but it also has the capability to write natural language grammar as well. ANTLR contain frameworks for compilers, recognizers and translators. It is implemented in Java but it can generate source code in Java, C, C++, C#, Objective C, Python and Ruby. ANTLR use EBNF (Extended Backus Naur Form) for the grammars, which is very formal way to describe the grammar. ANTLR provides a standard editor for grammar writing and generating lexer and parser. Till now this tool has been used for programming language's grammar writing but I am considering for natural language processing by writing natural language's grammar and generating lexer and parser to make the machine to understand it. ANTLR allows for generation of parsers, lexers, tree parsers and combined lexer parsers. Parsers can automatically generate Abstract Syntax Trees which can be further processed with tree parsers. ANTLR provides a single consistent notation for specifying lexers, parsers and tree parsers. This is in contrast with other parser/lexer generators and adds greatly to the tool's ease of use. By default ANTLR reads a grammar and generates a recognizer for the language defined by the grammar.

ANTLR has many belonging applications and opportunities to extensibilities. One of the biggest benefits is the grammar syntax; it is in EBNF form, which is a Meta syntax notation. Each EBNF rule has a left hand side (LHS) which gives the name of the rule and a right hand side (RHS) which gives the exact definition of the rule. Between the LHS and RHS there is the symbol ":" (colon), which separates the left from the right side and means "is defined as". One another benefit is the graphical grammar editor and debugger called ANTLRWorks, written by Jean Bovet and gives us the possibility to edit, visualize, interpret and debug any ANTLR grammar. It is based on a grammar editor with an interpreter for rapid prototyping and a language agnostic debugger for isolating grammar errors. ANTLRWorks also helps in eliminating grammar nondeterminisms by highlighting nondeterministic paths in the syntax diagram associated with a grammar. ANTLRWorks helps in making grammars more accessible to the average programmer by improve maintainability and readability of grammars and providing excellent grammar navigation and refactoring tools. To meet aforementioned goal of this research and development with respect to the implementation of an intelligent search by processing natural language, ANTLR can be used to take help in writing lexer and parser for own written natural language grammar.

## 4. Proposed Approach

Keeping eyes on above discussed major currently faced challenges of Product Data Management Systems, we can say, right now PDM community is in need of a very convincing and strong approach for its clients to win their confidence over the PDM Systems. Moreover PDM community also needs a new approach which can be very helpful in implementing the concepts of Product Data Management in the form of a web based software application capable of providing a user friendly graphical user interface which can also intelligently handle user's structured and unstructured natural language based requests for fast, optimized and efficient information retrieval or search mechanism

Targeting some of existing Product Data Management System development issues, proposed an approach, which was first conceptually modeled then converted in to implementation designs and which was then developed in the form of a prototype application. Proposed approach consists of four different modules i.e. Flexible GUI, NLP Search, Data Manager and Data Representer. Proposed approach is mainly for the development of a PDM system capable of providing a flexible web based graphical user interface, identifying user's structured and unstructured natural language based requests, processing natural language based user's requests to extract results from attached repositories, manage data in database management system and represent system outputted information as the result of user input in user's understandable format. Residing within the scope of this paper only discussing the NLP Search module in detail, as it is designed and implemented targeting the above mentioned unresolved issue of unintelligent search in PDM Systems.

## 5. NLP Search

Currently available data search mechanism in almost al PDM systems is not very efficient and based on old ways of searching information by entering the relevant information to the respective fields to search some specific information from attached repositories. As shown in Figure 5, if a user in need of some information, whether it is available in attached different networks sources or it is a CAD design or it is a document based information, in all the cases user has to spend some time in administrating the PDM system with the information he has about the thing which he is looking for e.g. to look for a document with specific type, date and author, user has to enter these relevant information entities in respective fields of search form and then system will check for the availability of that information. In case user does not know the complete or relevant information about the object he is looking for then it can be time consuming by increasing the manual efforts at user end. Targeting the problem of this unintelligent and old fashioned way of searching data in PDM Systems, I propose a natural language based search mechanism i.e. NLP Search for PDM System development.

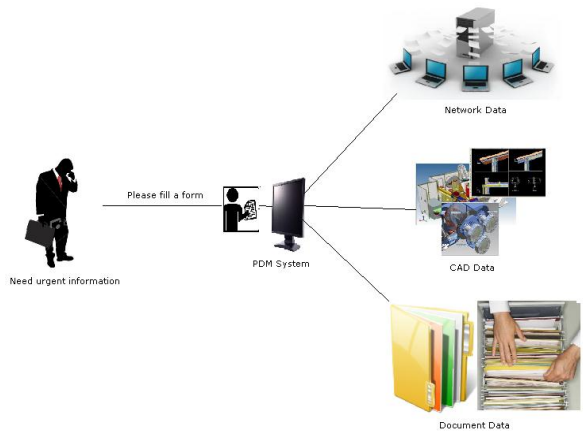

Fig. 5. Search in mechanism in PDM Systems

NLP Search is the second most important module of the proposed approach. This module is proposed by targeting the problem of unintelligent search in PDM Systems. The proposed job of this module is to take natural language based instructions from user to extract or search user needed information from attached repository. The proposed mechanism in this module is based on the extracted knowledge obtained as the result of conducted research in the field Language Technology and Semantic web.

To meet the aforementioned goal of a NLP search mechanism implementation, we need to propose a human machine communication system capable of translating user's natural language based instruction to machine understandable format. This data translation can be performed by structuring and categorization of data. To achieve this goal, a grammar is needed to be written based on the dictionary and rules of natural language used for user system communication. Residing within the scope of this research, a grammar is written which is based on small dictionary of keywords and set of rules. The designed grammatical view of the grammar is shown in figure 6. The grammatical view is designed with respect to English language grammar (rules) and consists of three parts i.e. A-Subject, B-Verb and C-Object, the simplest grammatical rule of used language i.e., "English".

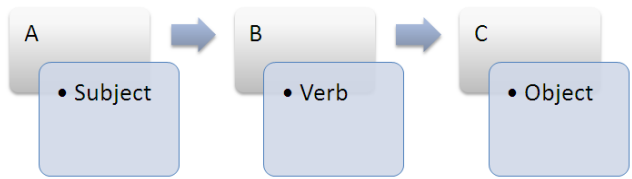

Fig. 6. Abstract View of Grammar

A is the subject; the set of tokens representing Grammatical Persons e.g. "I, We, He, she etc". B is the representation of Verbs and Helping verbs e.g. "is, need, want, look, give etc" and C is the Object; a set of tokens representing Nouns like "document, project, life cycle, person etc". These three parts most of the times combine to create a natural language based sentence e.g. "I am looking for PDM Systems" etc. Furthermore six different conditions are also introduced and used in this grammar i.e. Between, Euqal, Greater, Less, Greater Than, Less Than and With, as shown in Figure 7.

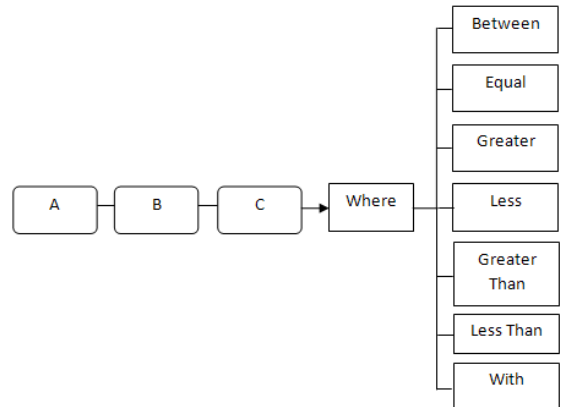

Fig. 7. Grammatical View with Conditions

Using this grammar four kinds of natural language based instructions can be processed i.e., Keyword based instructions, Short conditional objective statements, Simple conditional objective statements and Multiple Conditions based Objective statements.

- Keyword based instructions; user can simply enter any of these or similar words and can look for relevant information e.g. "PDM", "CAD", and "Documents" etc.
- Short conditional objective statements; where user is looking for some information by just combining some keywords e.g. "CAD Designs" and "PDM Documents" etc.
- Simple conditional objective statements; where user is looking for some information by writing simple unconditional defined natural language grammar based instructions e.g. "I am looking for PDM Document", "I need CAD Designs" etc.
- Multiple Conditions based Objective statements; where user is looking for some information by writing defined natural language grammar based instructions with more than one conditions e.g. "I am looking for PDM Documents where Document Type is doc", "Give me CAD designs of Car parts"

Following context and rules of the grammar, lexer and parser are also written. Lexer consists of the following sets of tokens i.e., Digits, Numbers, Subject, Verb, Object, Blanks and Conditions.
- Digits are the numbers from 0 to 9

- Numbers are the combinations of digits e.g. 123 etc.
- Subject is the set of tokens representing Grammatical Person
- Verbs is a set of tokens representing Verbs and Helping verbs
- Object is a set of tokens representing
- Nouns the possible actual objects
- Blanks are the empty spaces
- Conditions are the tokens representing conditional words e.g. where, between, Equal, And, these words can be used to make conditional statements like "I am looking for CAD where document equals to Screw", "We are looking for Project details between Date 01-09-08 and 01-09-09", "I want Document where Author equal to Michael" and "I need Product and Project Document" etc.

Whereas Parser consist of the following sets of five direct search rules i.e. astmt, bstmt, cstmt, stmt1, stmt2 and four conditional search rules i.e., condbt, condeq, condweq. Condeqbt have been created which are as follows;

- rule astmt represents only Subject from in process natural language based search queries like "I, We etc."
- rule bstmt represent only Verbs and Helping verbs from in process natural language based search queries like "need, look, give, am, are etc.",
- rule cstmt represent only Objcet in process natural language based search queries, etc "document, person , project etc.".
- rule stmt1 is the combination of Subject, Verbs and Helping Verbs and Object, to analyze natural language based search queries like "I am looking for CAD document" etc.
- rule stmt2 is the combination of Verbs and Helping Verbs and Object to analyze natural language based search queries like "give CAD desing" etc.
- rule condbt is the combination of Subject, Verbs and Helping Verbs, Object and Condition "between" e.g " I am looking for CAD Design between Number 100 and 200" etc.
- rule condeq is the combination of Subject, Verbs and Helping Verbs, Object and Condition "equal" e.g. " I need CAD with name MotorEngine and type BMP" etc.
- rule condweq is the combination of Subject, Verbs and Helping Verbs, Object and Conditions "with" and "equal" e.g. "I want Project with PDM name PDMDatabase" etc.
- rule condqbt is the combination of Subject, Verbs and Helping Verbs, Object and Conditions "with", "equal" and "between" e.g. "looking for a Project where PDMDatabase name  is between 2000 to 2009 Date" etc.

After having a grammar, the overall job of NLP Search module is divided into five steps .i.e., data reading, tokenization, parsing, semantic modeling and query generation, each step requires intensive effort in design and development. The main concept behind the organization of these five steps is to first read the user input natural language based instruction and to understand the semantic hidden in the context of natural language based instruction by lexing and parsing it. Then generate a query (SQL) to extract the desired results from attached repository. The implementation designs are constructed and discussed for prototype implementation of NLP Search using the concepts and technologies for Language Technology. The grammar is written using Antlr, lexer, parser and rest of the NLP Search module is developed by constructing implementation designs, and with the user of Java programming language.

## 6. Implementation Designs of Proposed Approach

To implement the proposed approach as a prototype (software) application, taking advantage from the observed information as the result of conducted research in the field of PDM System Development, I have designed a classical tier architecture consisting of three layers i.e., Presentation Layer, Business Logic and Database, as shown in figure 8.

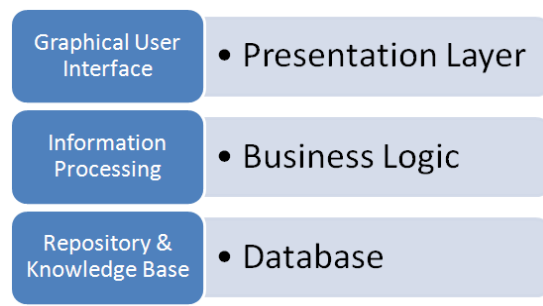

Fig.  8. Classical Three Tier Architecture

Presentation layer is the Graphical user interface of the prototype, carrying the jobs of user system communication. Business Logic is the information processing module of the prototype, carrying the jobs of transferring the user and system data between graphical user interface and database by implementing a communication system. Third layer is the Database, the main repository of the prototype, proposed to store, secure and manage data. This layer is the back end database management system, used to store and manage product attribute data and documentary information, as well as the relationships between data. This DBMS is usually a relational database system which provides complete functionalities to manage the product implemented using MySQL database management system.

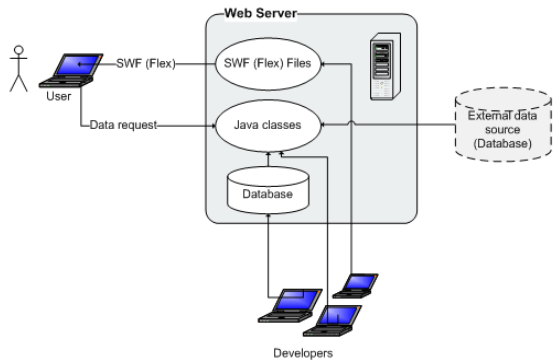

Fig. 9. Design Methodology

Following three tier application model of proposed approach, I have designed implementation methodology for the development of proposed prototype, as shown in Figure 9. The current version of proposed approach will be implanted with the use of Java (servlets and JSP) to handle user input, manage and retrieve data from the database. Tomcat is used as the main web server and middleware of the program. Users can access the web pages with the given URL and then can build graphical user interface or search the data after successful identity authorization. The data communication between three tiers is managed by Action Message Format (AMF) using the Simple Object Access Protocol (SOAP). AMF based client requests are delivered to the web server using Remote Procedure Call (RPC). The use of RPC allows presentation tier to directly access methods and classes at the server side. When data is request from user then a remote call is made from the user interface in the remote services' (via the server side includes) class members and the result is sent as an object of a Java class. A web browser is mainly needed to access the developed application with a user of a specified universal resource link (URL). User will send a request to the web server through Hypertext Transfer Protocol (HTTP), the web server will pass the request to the application components. These application components are implemented using servlet/JSP, designed to handle user request coming from web server with the use of java remote classes. Then used servlets or JSP classe talks to the database server, perform the data transactions and send the response to the client. To increase flexibility of graphical user interface at client end, the development of front end is performed using Flex Flex (Builder 3 IDE), earlier discussed in section section 3.2. Relational database is designed and implemented using MySQL 5.

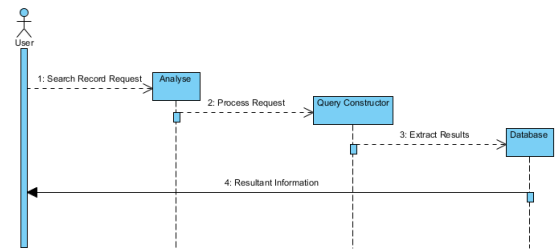

Fig. 10. System Sequence Design

As shown in figure 10, the System Sequence design consists of the three main components .i.e., Analyze, Query Constructor and Database. These three components perform certain jobs, the job of Analyze is to check if the it's a natural language based required from user to search record, then Query Constructor creates a SQL query to extract data from database and provide that to the user. Process and Model Information component first will store the information in to Repository, then will process the information by lexing, parsing and semantically modeling In the end Process and Model Information component will first save the information in Repository and then return the final system output to Graphical Interface.

## 7. NLP Search Prototype

This is the search module of developed Web Application. It can be accessed using Search Link from the links on the main. This prototype version is NLP Search, residing with in the limited scope, providing a search mechanism capable of processing natural language based user requests to extract desired information. The over all job of this module is explained in table 1 and shown in Figure 11 and 12.

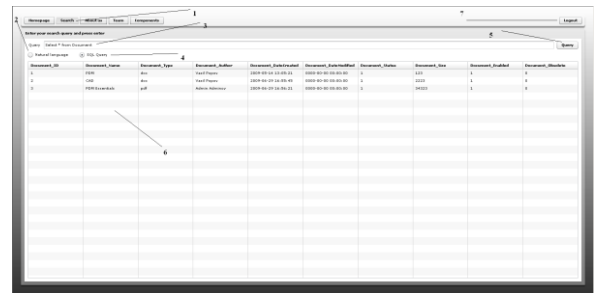

Fig. 11. I-SOAS Prototype- SQL based Search

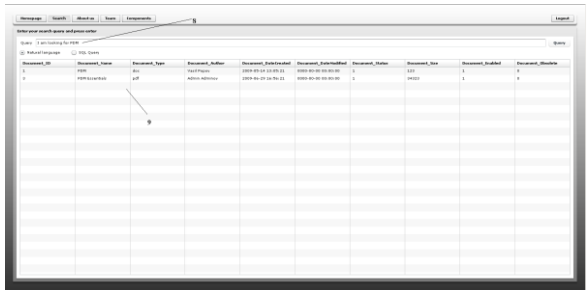

Fig. 12. I-SOAS Prototype- Natural Language based Search

## 8. Comparison with CIM and Windchill

Residing within the scope, meeting the goals of this research work and to evaluate the effectiveness of NLP Search module, some comparisons has been performed amongst developed Prototype, Windchill and CIM. During comparison earlier discussed four kinds of natural language search instructions i.e., Keyword based instructions, Short conditional objective statements, Simple conditional objective statements and Multiple Conditions based Objective statements, have been compared with the existing search module of CIM Database.

### 8.1. CIM and Prototype Comparison

In this section the search module of CIM Database has been compared with the search module of implemented prototype version. During comparison earlier discussed four kinds of natural language search instructions i.e., Keyword based instructions, Short conditional objective statements, Simple conditional objective statements and Multiple Conditions based Objective statements, have been compared with the existing search module of CIM Database.

#### 8.1.1. Keyword based instruction based comparison

A search was made with only one keyword

"PDM"

using both the search module of CIM Database as shown in Fig. 13 and prototype as shown in Fig. 14. As the result, both the applications performed accurately and the resultant information was presented by both systems, which proves that both the systems are equally efficient while making one keyword based search.

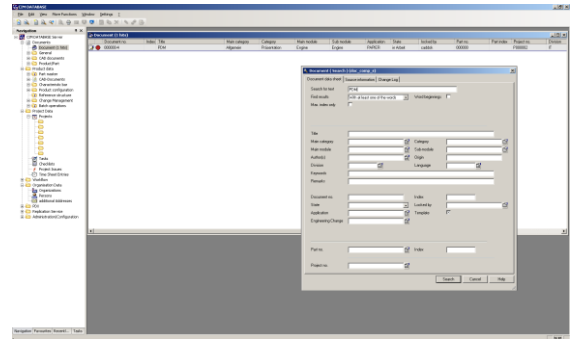

Fig. 13. CIM Database– Keyword based Search

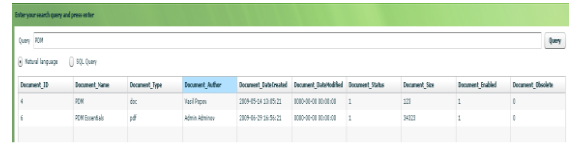

Fig. 14. Prototype – Keyword based Search.

#### 8.1.2. Short Objective Statement based comparison

A search was made with short objective statement,

"want PDM"

using both the search module of CIM Database as shown in Fig. 15and Prototype as shown in Fig.16. The resultant Fig. 15 of CIM Database clearly shows that the CIM Database was unable to find any output even when the result existed, moreover CIM Database search doesn't allow user to enter small alphabets because it is case sensitive. Whereas the resultant Fig. 16 of Prototype NLP Search module is presenting the obtained results. This comparison clearly proves that the text based search mechanism of CIM Database is unable to compile short objective statement whereas the Prototype NLP Search module successfully compiled short objective statement and presented the results.

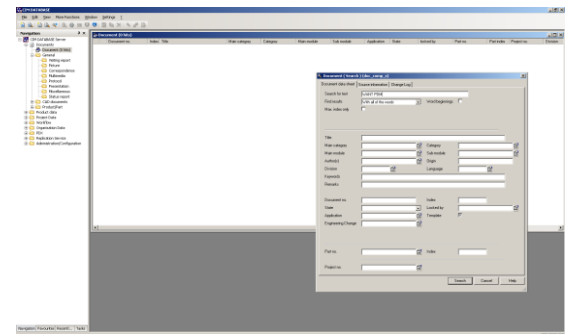

Fig. 15. CIM Database – Short Objective Statement based Search
.

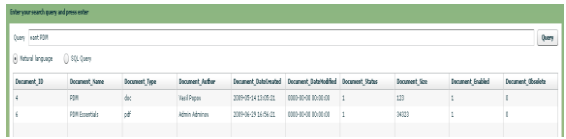

Fig. 16. Prototype– Short Objective Statement based Search.

### 8.1.3. Simple Objective Statement based comparison

A search was made with simple objective statement,

"I am looking for PDM"

using both the search module of CIM Database as shown in Fig. 17 and Prototype as shown in Fig. 18. The resultant Fig. 17 of CIM Database clearly shows that the CIM Database was unable to find any output even when the result existed, whereas the resultant Fig. 18 of Prototype NLP Search module is presenting the results. This comparison clearly proves that the text based search mechanism of CIM Database is unable to compile simple objective statement whereas the Prototype NLP Search module successfully compiled simple objective statement and presented the results.

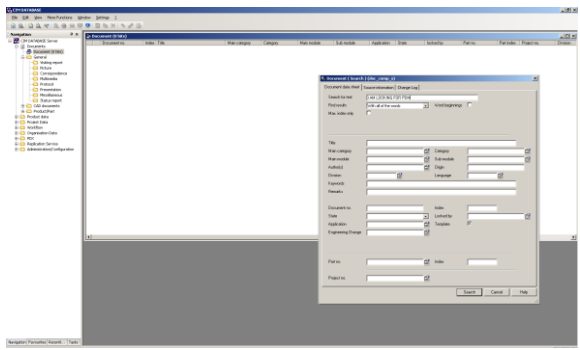

Fig. 17. CIM Database – Simple Objective Statement based Search.

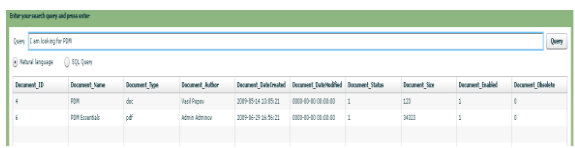

Fig. 18. Prototype – Simple Objective Statement based Search

### 8.1.4. 4. Multiple Conditions based Objective Statement based comparison

A search was made with multiple conditions based objective statement using both the search module of CIM Database as shown in Fig. 19 and Prototype NLP Search as shown in Fig. 20. This time the search module of both the applications was compared with the same kind of search methodology (natural language based full text search) like in previous comparisons. Because we have already concluded from previous comparisons that the natural language based full text search module of CIM Database is unable to compile short and simple objective statements to search the information, so it would be unwise to expect CIM Database search module to compile natural language based multiple conditions based objective statements.

So this time the provided search form based options were used to find some results by filling options with relevant information (conditional information). As shown in Fig. 19 that there are several different options on the Document search form of CIM Database to find out the information about a document e.g. with some specific type or category etc. All required is to fill the search form by entering or selecting required information e.g. type as shown in Fig. 19. Whereas as shown in Fig. 20, in case of Prototype NLP Search module you just need to enter the natural language based search query .e.g.

"I am looking for PDM where Document Type is doc"

and the relevant result will appear. In this comparison both the search modules of the applications bring results but the major difference is not of the results but of the ease in use of the software. In case of CIM Database a user at first needs to open relevant search form e.g. document search or product search etc. and then has to enter relevant information in respective text boxes and then might also need to open some other forms and select some other information e.g. category etc. and then he can make a search. But in case of Prototype NLP Search instead of providing all this information user only needs to enter a simple natural language based multi conditional statement, which not only provides the ease in use but also reduces the time and effort of the user in searching required information.

As earlier mentioned that the Prototype NLP Search module also allows a professional database user to enter the SQL based query of his own choice to search the results, Fig. 21 presents the obtained results using SQL query

"select * from document where Document_Type = "doc" and Document_Name = "PDM""

The resultant information not only satisfies the user but also confirms the result of above used natural language based multiple condition based object statement. This search with Multiple Conditions based Objective statement comparison proves that the required effort needed to search multi conditional information is very little in Prototype NLP Search module as compared to the CIM Database search module.

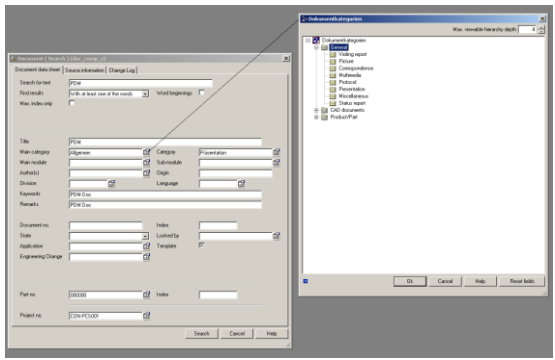

Fig. 19. CIM Database – Multiple Conditions Search

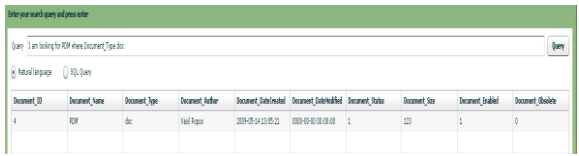

Fig. 20. Prototype – Multiple Conditions based Search

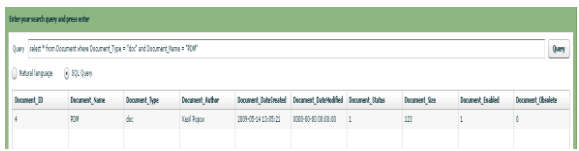

Fig. 21. Prototype – SQL Search Query based Conditional Search

### 8.2. Windchill and Prototype Comparison

In this section of paper the search module of Windchill has been compared with the search module of implemented prototype version of Prototype NLP Search. Likewise CIM Database's comparison with developed Prototype, I have compared four kinds of natural language based search instructions i.e., Keyword based instructions, Short conditional objective statements, Simple conditional objective statements and Multiple Conditions based Objective statements, with the existing search module of Windchill and Prototype NLP Search.

### 8.2.1. 1. Keyword based instruction based comparison

A search was made with only one keyword,

"CAD"

using both the search module of Windchill as shown in Fig. 22 and Fig. 23, and Prototype NLP Search as shown in Fig. 24. As show in figure Fig. 22 when the keyword CAD is entered in search text box of the main Home page of the Windchill and searched, a new page "Advanced Search" is appeared as shown in Fig. 23. The following result shown in Fig. 23 is obtained when CAD keyword is searched using advanced search page. Likewise as shown in the Fig. 24, when the keyword CAD is searched using Prototype NLP Search module then the CAD information is obtained. As the result both the applications performed accurately and the resultant information was presented by both systems, which proves that both the systems are equally efficient while making one keyword based search.

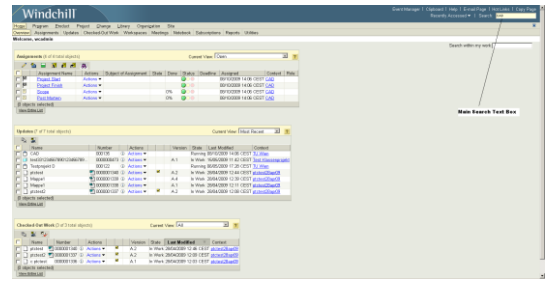

Fig. 22. Windchill Search –Main

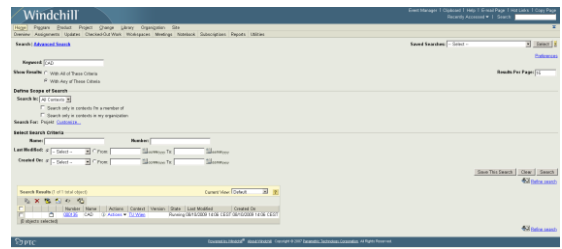

Fig. 23. Windchill – keyword based Search

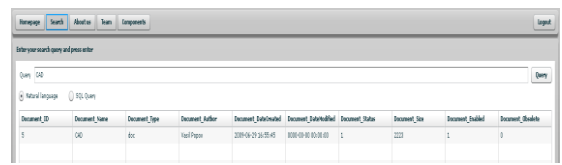

Fig. 24. Prototype – Keyword based Search

### 8.2.2. Short Objective Statement based comparison

A search was made with short objective statement,

"need CAD"

using both the search module of Windchill as shown in Fig. 25 and Prototype NLP Search as shown in Fig. 26. The resultant Fig. 25 of Windchill clearly shows that likewise CIM Database Windchill is also unable to find any output even when the result existed. Whereas the Fig. 26 of Prototype NLP Search module presents the results. This comparison clearly proves that the text based search mechanism of Windchill is unable to compile short objective statement whereas the Prototype NLP Search module successfully compiles short objective statement and gives the results.

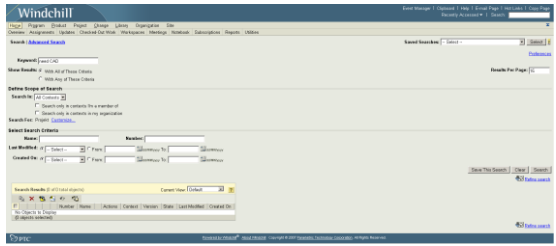

Fig. 25. Windchill – Short Objective Statements

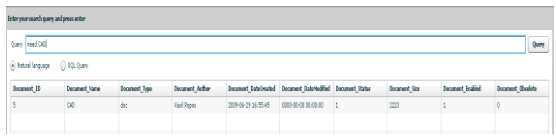

Fig. 26. Prototype Short Objective Statements based Search

### 8.2.3. Simple Objective Statement based comparison

A search was made with simple objective statement,

"He is looking for CAD"

using both the search module of Windchill as shown in Fig. 27 and Prototype NLP Search as shown in Fig. 28. The resultant Fig. 27 of Windchill clearly shows that Windchill is again unable to find any output using simple objective statement with using both the options of text i.e., With All of These Criteria and With Any of These Criteria, even when the result existed whereas the resultant Fig. 28 of Prototype NLP Search module presents the results. This comparison clearly proves that the text based search mechanism of Windchill is unable to compile simple objective statement whereas the Prototype NLP Search module successfully compiles simple objective statement and presents the results.

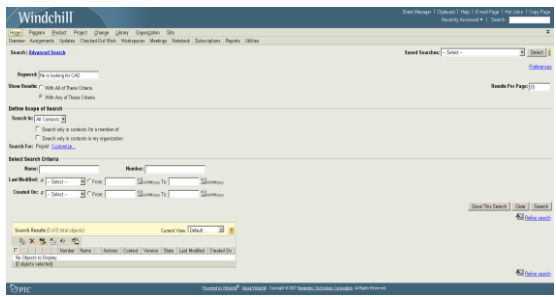

Fig. 27. Windchill – Simple Objective Statement based Search

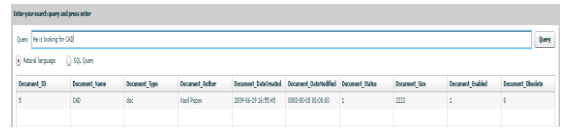

Fig. 28. Prototype NLP Search Query – Simple objective statement

### 8.2.4. Multiple Conditions based Objective Statement based comparison

A search was made with multiple conditions based objective statement using both the search modules of Windchill as shown Fig. 29 and Prototype NLP Search as shown in Fig. 31. Like the conditions based objective statement comparison of CIM Database and Prototype NLP Search, here again this time I am not comparing both application's the search module with same kind of search methodology like I did in previous comparisons. As we have already concluded from previous comparisons that the Search module of Windchill is unable to compile short and simple objective statements to search the information, so this time I am not comparing search statements. But I am using search form based options to find some results by filling options with relevant information (conditional information).

As shown in Fig. 29 that there are some options on the advanced search form of Windchill to find out the information. To obtain any information by making multiple conditions based search, user is required to first fill some form based options, and if options of user interest are not available in the search form then user must click on "Customized" option to first choose the relevant options to create appropriate search form, which is again a hectic task. For example user is interested in finding out a document named CAD with some specific type e.g. "doc" and "pdf". As shown in Fig.29, user starts the search using word name CAD but the resultant information is based on CAD project. If user is not satisfied with this output then he needs to first customize the search page as shown in Fig. 29. Moreover by looking at this search output a user might think that there is no such document exists with the name CAD, but as clearly shown in Fig. 30 that there is a document with name CAD in the system. Whereas in case of Prototype NLP Search module user just needs to enter the natural language based search query .e.g.

"She need PDM with Document Type doc and Pdf"

As shown in Fig. 31, and the relevant results are presented. This search with Multiple Conditions based Objective statement comparison proves that the required effort needed to search some multi conditional information is very little in Prototype

NLP Search module as compared to the Windchill search module. Moreover in case of Windchill, for making different kinds of search like searching documents, projects, products etc. each time advanced search form will appear, which requires user to train himself for performing different kinds of search consuming time and effort where as in case of Prototype NLP Search module, user does not need to learn to use multiple options based search form to carry out different searches, moreover he can use simple natural language based queries to search required information.

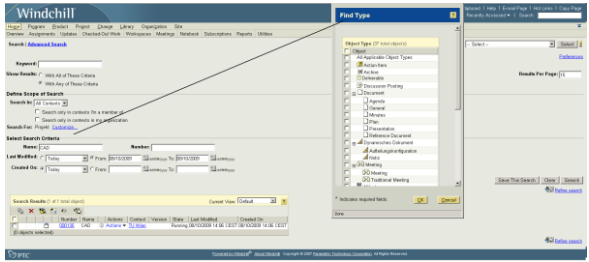

Fig. 29. Windchill – Multiple Conditions based Search

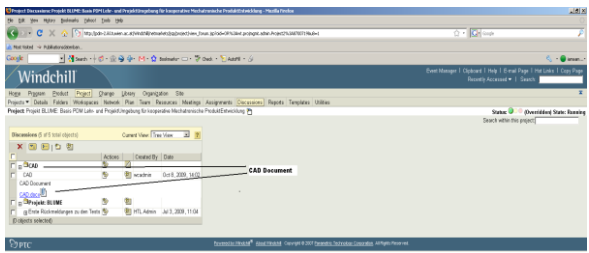

Fig. 30. Windchill – CAD Document

Fig. 31. Prototype – Multiple Conditions based Search

## 9. Limitations

We have proposed and written a new grammar for natural language processor implementation for PDM Systems but due to the limited scope of this research and development work, the word length of proposed dictionary for lexer is small and the number of rules designed for parser are also a few. At the moment only seven different set of tokens for lexer i.e. Digits, Numbers, Subject, Verb, Objects, Blanks and Conditions, and five rules are defined for parser implementation.

## 10. Conclusion

Targeting the challenge of proposition of natural language based search, a thorough research has been conducted in the field of Language Technology and findings are presented in this paper. Taking help from observed information from conducted research in respective field and using person research and development experience I have proposed an approach. We have designed conceptual and implementation designs of proposed approach and implemented it using some software tools and technologies of present time i.e. Flex, Java, Antlr, MySQL, and presented developed prototype solutions. In the end concluding the research and development efforts, I can say that the proposed approach can put some values in enhancing PDM System development process. The inclusive implementation of the newly proposed idea in PDM System development can put some values in increasing the market values of PDM Systems by increasing its acceptability in industry by improving its use amongst managerial, technical and office staff.

## 11. Acknowledgments

I am thankful to University of Wuerzburg Germany and Vienna University of Technology Austria for giving me the opportunity to keep working on this research project. I am thankful to Prof. Dr. Detlef Gerhard for his supervision during this research and pay my gratitude to Prof. Dr. Thomas Dandekar for his generous support. I also thanks to my beloved wife Mrs. Saman Zeeshan (Majeed) for her support during this research, development and technical documentation.

**Author Bibliography**

Zeeshan Ahmed is a Software Research Engineer, presently working in the Department of Bioinformatics, Biocenter, University of Wuerzburg Germany. He has on record more than 12 years of University Education and more than 8 years of professional experience of working within different multinational organizations in the field of Computer Science with emphasis on software engineering of product line architecture based artificially intelligent systems. He also has more than 4 years experience of teaching as lecturer and supervising research thesis to graduate and undergraduate students in different institutes and universities.

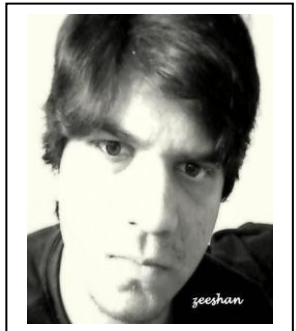